# Introducing open boundary conditions in modeling nonperiodic materials and interfaces: the impact of the periodicity assumption

James Charles[1,*], Sabre Kais[2,3], and Tillmann Kubis[1,4,5,6]

[1]SCHOOL OF ELECTRICAL AND COMPUTER ENGINEERING, PURDUE UNIVERSITY, WEST LAFAYETTE, INDIANA 47907, USA

[2]DEPARTMENT OF PHYSICS AND ASTRONOMY, PURDUE UNIVERSITY, WEST LAFAYETTE, IN, USA

[3]DEPARTMENT OF CHEMISTRY, PURDUE UNIVERSITY, WEST LAFAYETTE, IN, USA

[4]NETWORK FOR COMPUTATIONAL NANOTECHNOLOGY, PURDUE UNIVERSITY, WEST LAFAYETTE, INDIANA 47907, USA

[5]CENTER FOR PREDICTIVE MATERIALS AND DEVICES, PURDUE UNIVERSITY, WEST LAFAYETTE, INDIANA 47907, USA

[6]PURDUE INSTITUTE OF INFLAMMATION, IMMUNOLOGY AND INFECTIOUS DISEASE, WEST LAFAYETTE, INDIANA 47907, USA

*KEYWORDS Quantum modeling, Green's function method, Open boundary method, Defects, Interfaces.*

**ABSTRACT:** Simulations are essential to accelerate the discovery of new materials and to gain full understanding of known ones. Although hard to realize experimentally, periodic boundary conditions are omnipresent in material simulations. In this work, we introduce ROBIN (recursive open boundary and interfaces), the first method allowing open boundary conditions in material and interface modeling. The computational costs are limited to solving quantum properties in a focus area which allows explicitly discretizing millions of atoms in real space and to consider virtually any type of environment (be it periodic, regular, or random). The impact of the periodicity assumption is assessed in detail with silicon dopants in graphene. Graphene was confirmed to produce a band gap with periodic substitution of 3% carbon with silicon in agreement with published periodic boundary condition calculations. Instead, 3% randomly distributed silicon in graphene only shifts the energy spectrum. The predicted shift agrees quantitatively with published experimental data. Key insight of this assessment is, assuming periodicity elevates a small perturbation of a periodic cell into a strong impact on the material property prediction. Periodic boundary conditions can be applied on truly periodic systems only. More general systems should apply an open boundary method for reliable predictions.

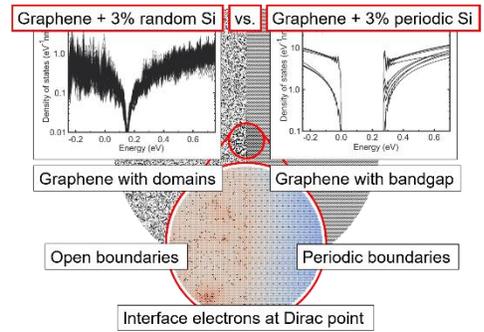

Computer aided material predictions represent the first-step of many new material discoveries[1–3]. Material simulations can power machine learning searches for new materials with specific properties[4–6]. However, modeling experimental reality with wide-spread idealized, periodic boundary conditions[7,8] is prone to artifacts: Irregular interfaces, impurities, cracks and dislocations are not compatible with idealized conditions. A common approach to limit artificial periodicity effects is to make the repeating unit cell as large as numerically feasible and apply various correction algorithms[9–12].

Instead, we introduce the Recursive Open Boundary and INterfaces (ROBIN) method that can handle arbitrary geometries and atom distributions and does not need any periodicity assumption. It is based on the nonequilibrium Green's function method (NEGF). The NEGF method had been applied on charge[13,14], spin[15,16] and heat[17,18] transport in open nanodevices. The ROBIN extension of NEGF models materials in infinitely extended real space and supports regular and irregular systems. We verify the ROBIN method in 2D and 3D crystalline systems. To assess the impact of periodic boundary conditions on material property predictions, an as-simple-as-possible, but experimentally realized system was chosen: Calculations of graphene confirm recent work[19] that periodically distributed silicon impurities can open bandgaps. In stark contrast and presumably closer to any experiment, random distributions of the same amount of silicon are shown to give no band gaps, but to form domains and to linearly shift the band structure. The predicted shift quantitatively agrees with experimental data of Ref.19. The findings of ROBIN are analyzed in detail and show that periodic boundary conditions can elevate otherwise small perturbations to systematic changes of material properties.

So far, all models for quantum electronic material properties are based on Hermitian Hamiltonian operators (H) that represent either periodic or finite sized systems[20]. The boundaries of closed systems yield confinement effects and system size dependent resonances that can interfere with the actual material properties. Models with periodic boundary conditions require numerically hard to achieve unit cell sizes to avoid artificial long-distance coupling between repeating simulation domain features[21]. To lift some of the numerical limitations of periodic simulations, various correction methods have been introduced[11,22,23]. The k-space sampling required for periodic boundary simulations represents additional numerical challenges[20]. Modeling systems with long distance effects such as Moiré lattices, systems with irregularities such as alloys and systems with inhomogeneous fields or strain are notoriously difficult to handle with Hermitian Hamiltonian operators.

In the NEGF method, the electronic density of states (DOS) equals the imaginary part of the retarded Green's function's ($G^R$) diagonal. $G^R$ is solved in the Dyson equation which reads in operator form $G^R = (E - H_C - \Sigma^R)^{-1}$, with the electronic energy E, and the retarded self-energy $\Sigma^R$ [24]. The Hermitian Hamiltonian $H_C$ represents the electrons in the finite, central area C. We set C to be a sphere for three-dimensional and a circle for two-dimensional systems. However, any other space-appropriate shapes are possible, too. Electrons are modeled in the effective mass approximation[25] when the ROBIN method is verified against analytical DOS of parabolic dispersions in 2D and 3D. In case of graphene, electrons are given in single-orbital atomistic tight binding ($E_{Pz,C} = 0$, $V_{PP\sigma,C} = 0$, $V_{PP\pi} = -3$ eV, following the nomenclature of Ref.26) on the native graphene lattice. Silicon atoms in graphene are modeled with graphene parameters and an onsite energy of $E_{Pz,Si} = 4.75$ eV to reproduce the band gap of 3% periodically distributed Si in graphene predicted with DFT in Ref.19. Note that many other electronic representations, such as plane waves[27], maximally localized Wannier functions[28,29] or localized atomic orbitals[30,31] have been applied in NEGF before. Devices modeled in NEGF covered 1D, 2D and 3D symmetries, ranging from molecular junctions[32] up to micrometer long resistors[33].

The retarded self-energy $\Sigma^R$ is the key element that distinguishes NEGF from closed-system models: It is the non-Hermitian operator in the inverse $G^R$ that represents the interaction of electrons in C with the surrounding of C at the contact interface between the two regions[34]. $\Sigma^R$ allows electrons to enter and leave C at the contact and then to propagate to infinite distance to C. The imaginary part of $\Sigma^R$ is inverse proportional to the electronic lifetime in C (i.e. the "dwelling-in-C-time")[35].

Most NEGF applications require the surrounding "behind" the contact to form a homogeneous lead and in particular to have a well-defined 1D transport direction. A few exceptions to this limitation can be found for quantum cascade systems[13,36] and recent transistor predictions[37]. Reference 37 in particular allowed for the lead cross section size to grow infinitely with increasing distance to the contact and to host random atom distributions.

The ROBIN method expands the contact self-energy method of Ref. 37 by considering the total interface between C and the surrounding as the contact area. The conceptual difference to Ref.37 is the fact that only one contact self-energy describes the complete environment. Following Ref.37, the non-Hermitian $\Sigma^R$ is solved as a product of the non-Hermitian surface retarded Green's function of the 2D or 3D surrounding of C with the Hermitian Hamiltonian operators of atoms in C coupling with atoms in the surrounding. Thereby, the environment atoms are discretized explicitly. A complex absorbing potential (CAP) is added to the environmental atoms' on-site energies[38]. Similar to Ref.37, the CAP vanishes at the edges of C and grows smoothly with increasing distance to C [39]. The CAP is critical to ensure efficient convergence of the results in C with the range of explicitly discretized surrounding atoms.

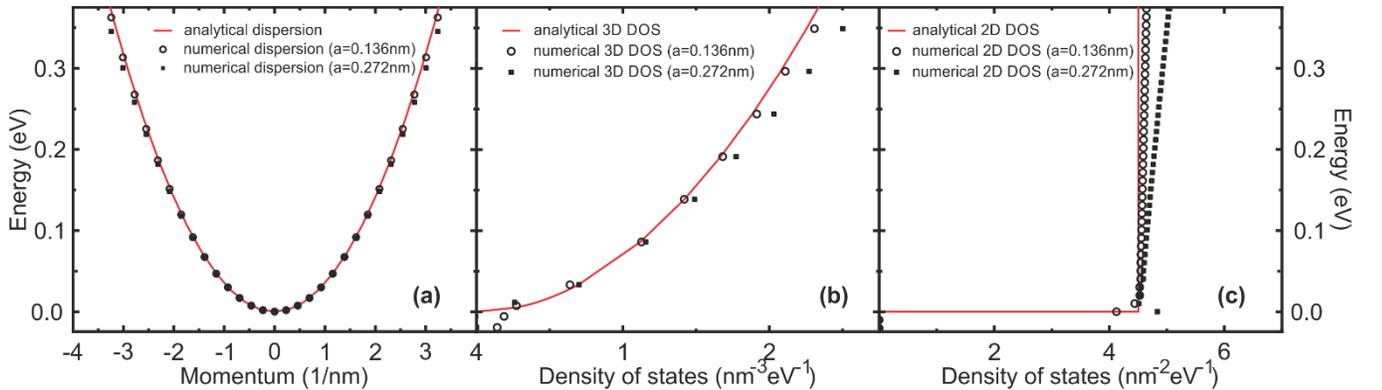

Figure 1 Verification of the ROBIN method against analytical results: (a) The analytical dispersion of effective mass electrons (line) and the numerical dispersion of periodic Hamiltonian operators discretized in a real space mesh of 0.136nm (circles) and 0.272nm (dots) mesh point distance are known to deviate more with higher kinetic energies. Similarly, the numerical density of states of open

system simulations with ROBIN (symbols) deviates from the analytical one (lines) with higher energies and larger mesh constants, both in 3D (b) and 2D (c). Otherwise, all results of the ROBIN method resemble the expected analytical data very well.

The numerical costs solving for the retarded Green's functions is the largest challenge of the ROBIN method. Therefore, all retarded Green's functions are solved recursively[40,41] to limit the required peak memory and to allow for explicit consideration of up to 3 million atoms in this work. Many publications[42–46] and online lectures[47,48] on recursive Green's functions describe the method in high detail. Details of the CAP method are discussed in Refs.37 and 49. All ROBIN calculations have been performed on 10 nodes of the Brown cluster of the Rosen Center for Advanced Computing at Purdue University.

Since all density of states results of open system calculations come with a continuous DOS, smoothing spectral results as needed in Hermitian models is obsolete here[50–52]. Although this work covers only electronic examples, the presented method applies to any system with discretizable equations of motion including e.g. lattice vibrations in dynamic matrix descriptions.

Figures 1 verify the ROBIN method for electronic material property predictions. Figure 1(a) is a reminder of the electronic dispersion resulting of electronic Hamiltonian operators of silicon conduction band electrons ($m^*=1.08m_0$) discretized in real space and solved with periodic boundary conditions. Note this is the only periodic-boundary system result, while all remaining results apply the ROBIN method of open boundaries. Deviations from the analytical parabolic dispersion become smaller with decreasing kinetic energies and finer mesh spacing[53]. Accordingly, with finer real space meshes and smaller kinetic energies the DOS of the ROBIN method in 3D (Fig. 1(b)) and 2D (Fig. 1(c)) agree better with the respective analytical DOS, i.e. the square root function in 3D and the constant DOS in 2D.

Similar to the Si nanowire calculations in Ref. 19, the convergence of $\Sigma^R$ close to band edges is more demanding and small deviations from the analytical DOS can be observed there. Better convergence further reduces the DOS deviation at the band edge.

This convergence also determines the quality of the predicted DOS at the Dirac point of graphene. Figure 2 shows the average DOS of graphene electrons solved in graphene discs of varying diameters. The center region C is chosen to be a disc of 1nm diameter for all results in Fig.2.

All remaining carbon atoms are included as part of the environment of C within the ROBIN method. In this way, the largest disc size considered in Fig. 2 is 200nm diameter which includes more than 1 million discretized carbon atoms in total. The average DOS in Fig.2 converges well to the linear dispersion of graphene with increasing lead size. Simultaneously, the standard deviation of the DOS of each considered atom in C vs. the depicted average value reduces, too. The maximum of this standard deviation for all considered energies in Fig. 2 is $1.2\times10^{-4}$ eV$^{-1}$nm$^{-2}$ (80nm), $1.5\times10^{-5}$ eV$^{-1}$nm$^{-2}$ (140nm), and $6.3\times10^{-7}$ eV$^{-1}$nm$^{-2}$ (200nm), respectively.

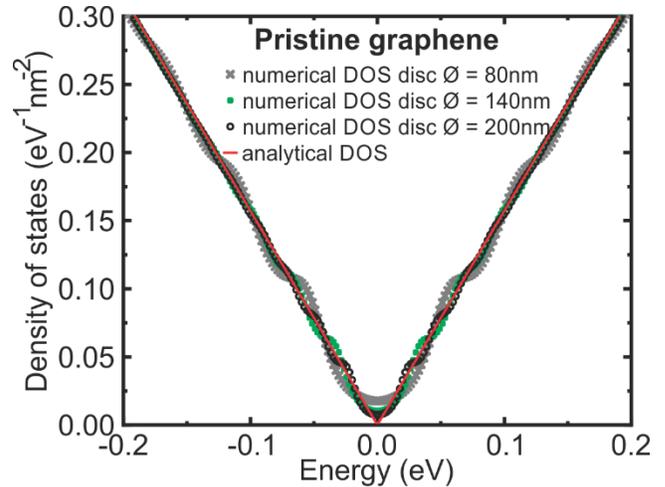

Figure 2 Verification of the ROBIN method against analytical results: The numerical density of states resulting of the ROBIN method (symbols) of graphene discs agree better with the analytical density of states (line) with larger discretized disc diameter.

In Ref. 19, a 3% concentration of periodically distributed silicon atoms in graphene was analyzed with density functional theory calculations and periodic boundary conditions. It was predicted that the addition of the silicon atoms opens a bandgap of 0.28eV in graphene. This finding can be reproduced with the ROBIN method in empirical tight binding: All Si atoms are considered periodically distributed in the graphene disc. Silicon parameters are approximated with graphene parameters and an additional onsite energy of 4.75eV. Given the unit cell is larger with the periodic Si than in the case of pristine graphene (see Fig. 3), the convergence of the DOS w.r.t the disc diameter is numerically more challenging. This can be seen in the slowly decaying beating pattern in Fig. 4 (a). Even a disc diameter of 320nm with more than 3 million discretized atoms still shows a small beating in the resulting DOS around the band gap. Figure 4(a) shows the electronic DOS of each of 282 atoms of a 3nm center area of two different graphene discs (200nm and 320nm diameter) solved with the ROBIN method.

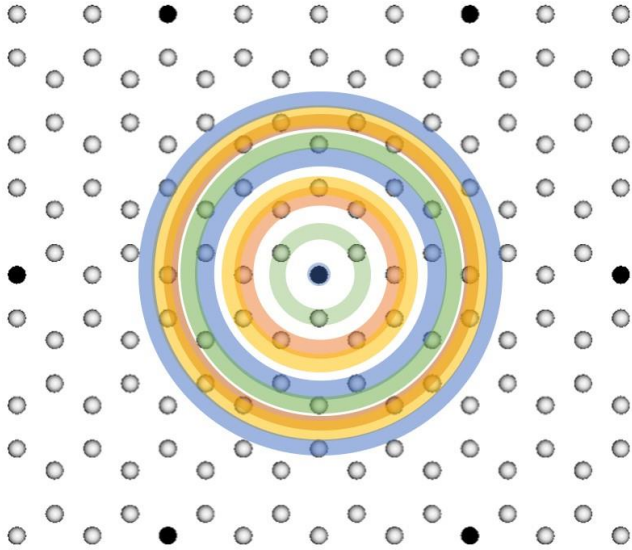

Figure 3 Schematic of carbon (white) and silicon (black) atoms in graphene with 3% periodically distributed silicon. 9 different atom types are in each unit cell: one silicon atom and carbon atoms in 8 different distances to the central silicon, highlighted by 8 semi-transparent rings around the center Si circle.

The periodic distribution of carbon (white) and 3% silicon (black) atoms is shown in Fig. 3. The addition of silicon atoms increases the graphene unit cell to 32 atoms that fall into 9 different chemical categories: 1 silicon atom and 8 graphene atoms in 8 different distances to the silicon one (see Fig.3). Accordingly, a ROBIN prediction of the atom resolved DOS of graphene with 3% periodically distributed Si yields 9 different DOS lines – as shown in Fig. 4(a). Note that Fig. 4(a) actually shows 282 individual DOS lines for each of the 282 atoms in the 3nm center region. Good convergence of the contact self-energy makes them virtually identical to DOS lines of atoms with the same chemical environment (see Fig. 4 (b) for a zoom-in).

The DOS changes significantly when the 3% silicon atoms are randomly distributed (see Fig.4 (c)). The 282 local DOS lines of each atom in the center region C differ depending on their respective local atomic environment. The ensemble of atomic DOS lines maintains a Dirac point at about $\Delta E=0.147eV$ above the Dirac point of pristine graphene. Note that $\Delta E$ scales approximately linearly with the % fraction of randomly distributed Si atoms in graphene as can be seen in Fig.4(c) for the 1% and 2% Si cases. For comparison, Fig. 4(c) also shows the analytical DOS of pristine graphene.

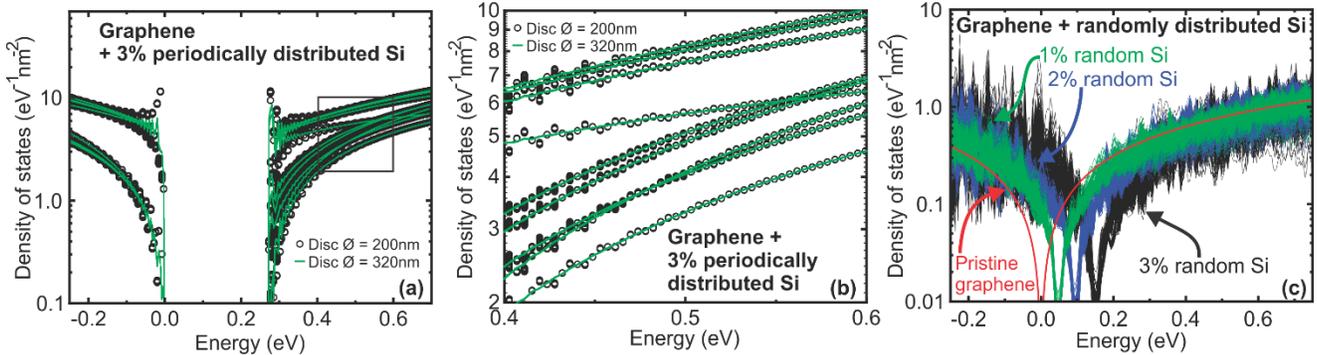

Figure 4(a) The density of states of graphene with 3% periodically distributed silicon solved with the ROBIN method reproduces the 0.28eV band gap of Ref.19 when the on-site energy of Si is chosen as 4.75eV. The large unit cell of 3% Si in graphene burdens the numerical convergence w.r.t the disc diameter. 200nm (symbols) and to lesser extend the 320nm (lines) disc diameter show incomplete convergence near the band gap. (b) Zoom-in into the boxed region in Fig. 4(a). The 282 individual atoms of the calculation in Fig.4(a) fall into 9 distinct groups of DOS lines – corresponding to the 9 different atom types shown in the Fig.3. (c) The DOS solved in the ROBIN method of randomly distributed Si atoms in graphene does not show a bandgap. Instead, increasing Si content shifts the DOS to higher energies by about 47meV per Si-percentage (i.e. about 1% of the assumed onsite energy difference of carbon and silicon atoms). The red line shows the analytical DOS of pristine graphene for comparison.

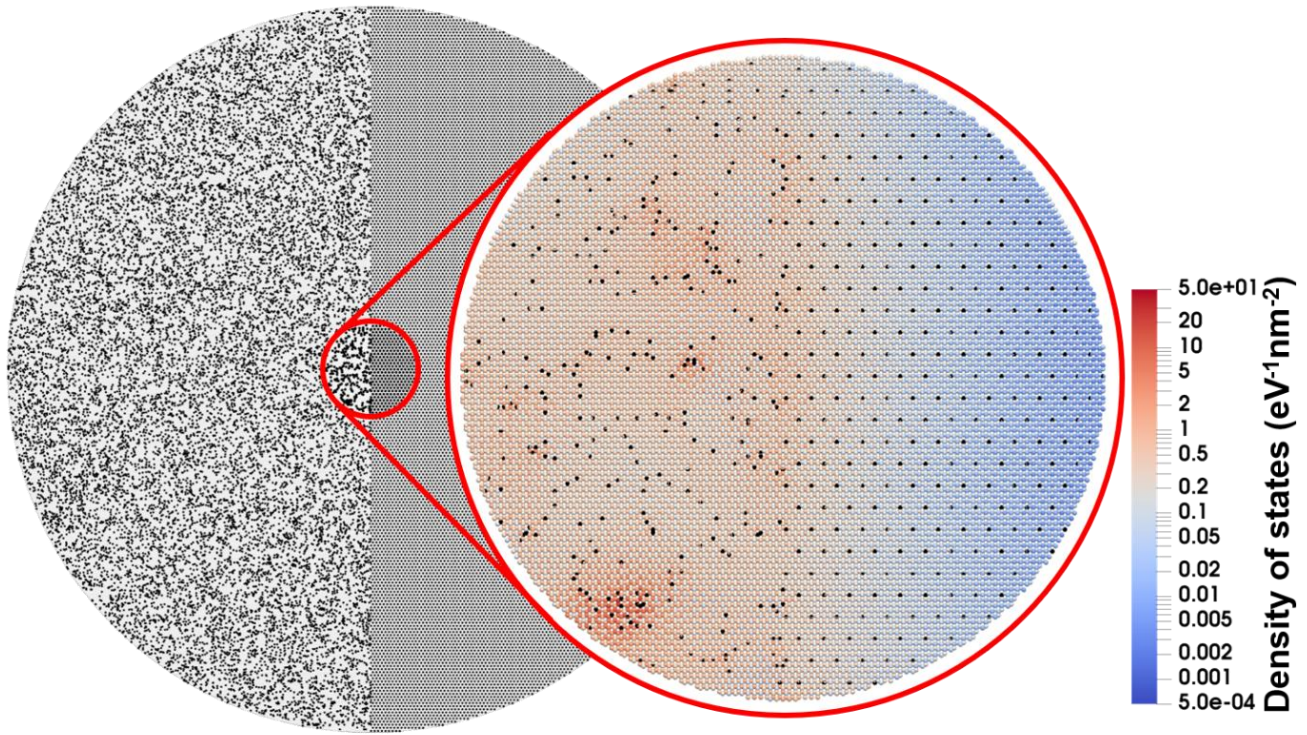

Figure 5(left) 200nm disc of graphene (carbon atoms are white) with 3% Si atoms (black) distributed randomly on the left, and periodically on the right half of the disc. (right) Electronic density of states of the center 25 nm of the 200nm graphene disc solved with open boundary conditions at 10meV above the Dirac point of pristine graphene. Carbon atoms are colored according to the electronic DOS, silicon atoms are black. The electronic DOS shows domain formation in the left half and electronic tunneling into the right half of the disc.

Adding only 1%, 2%, or 3% silicon should perturb graphene due to the linear response regime only. Indeed, the ROBIN results in Fig. 4 (c) for this amount of randomly distributed Si show only a linear shift of the Dirac cone. Periodic boundary conditions of the same small amount of Si atoms give a dramatic change to the graphene bandstructure, resembling effectively a new material. In other words, applying periodic boundary conditions elevates otherwise small perturbations to systematic material property changes. Therefore, periodic boundary conditions should only be applied to truly periodic systems. In the experiments of Ref. 19, 3% randomly distributed Si in graphene yielded a shift of the electronic work function by 0.13eV. The predicted shift of 147meV in Fig. 4 (c) agrees quantitatively with that observation given the experimental Si concentration uncertainty of Ref. 19 (2.7% - 4.5%).

To illustrate the DOS difference of periodically and randomly distributed silicon atoms in graphene, Fig. 5 shows open system results of the center 25nm of a 200nm diameter graphene disc with 3% silicon atoms distributed randomly on the left half and periodically on the right half of the disc. The contour shows the position resolved DOS at the energy of 10meV above the Dirac point of pristine graphene. The black spheres indicate the position of Si atoms. Depending on local Si atom distributions, electrons on the left face pockets of high DOS. Whereas all the DOS decays in the right due to the bandgap opened by the periodically distributed Si.

Substituting atoms periodically is a remarkably difficult experimental task especially if single substitutions are considered. We expect random distributions to resemble the experimental reality much more closely. Given the stark contrast in electronic properties of periodic vs. random distributions, materials with periodic substitutions should be considered fully distinct from the original pristine host material. This applies to substituting with other than Si atom kinds[54,55] as well as other host materials than graphene.

In conclusion, this work introduces the ROBIN method to predict 2D and 3D materials in arbitrary, regular, and irregular atomic compositions. Green's functions are solved recursively to explicitly discretize millions of atoms within the memory limitations of typical state of the art hardware. When applied on silicon atoms distributed in graphene, the method reveals a significant difference in the electronic properties of periodic vs. randomly distributed Si atoms in graphene. The calculations confirm periodically distributed Si atoms form bandgaps in graphene, but the same amount of randomly distributed Si atoms forms domains in the electronic DOS and shifts the graphene DOS in energy. The results show that applying periodic boundary conditions can elevate small perturbations to massively influence material property predictions.

It is worth to mention the ROBIN method can be applied on systems with random alloys, single defects and interfaces. Systems involving different physical phases (e.g. heterogeneous catalysis[56], emulsions[57], melting solids, microdroplet chemistry[58], etc.) are conceptionally equivalent to the situation in Figure 5. To illustrate the generality of the ROBIN method, Figs. 6 shows the electronic DOS at the energy of the graphene Dirac point in twisted bilayer graphene with a twist angle of 5 degrees (a) and 30 degrees (b), respectively. The ROBIN method is independent of whether the system is irregular, periodic or quasicrystalline. The data in both Figs. 6 were solved with the same numerical effort and the same simulation settings. The expected periodicity and quasicrystalline behavior is reproduced in both cases[59–61].

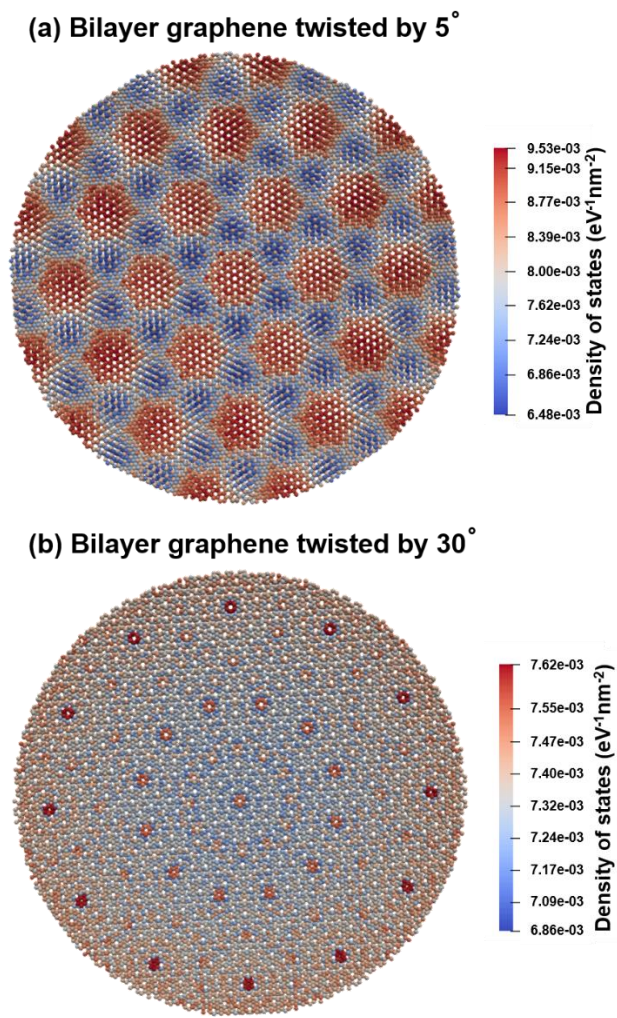

Figure 6 Electronic DOS at the energy of the graphene Dirac point in twisted bilayer graphene with a twist angle of 5 degrees (a) and 30 degrees (b). Bilayer graphene parameters of NEMO5's[14] database were altered with Harrison scaling for the electronic Hamiltonian. The results of both angles were solved with the same number of explicitly discretized atoms, independent of the resulting DOS distribution.

The applicability of ROBIN on 3D materials is exemplified in Figs. 7 as they show the electronic DOS of a spherical Ge cluster embedded in Si and solved with ROBIN in sp3d5s* atomistic tight binding representation[62]. The cluster size is chosen to be 2nm diameter common in SiGe alloys[63]. The electronic DOS at 50meV above the Si band edge has two maxima in the Ge cluster, but leaks significantly into Si.

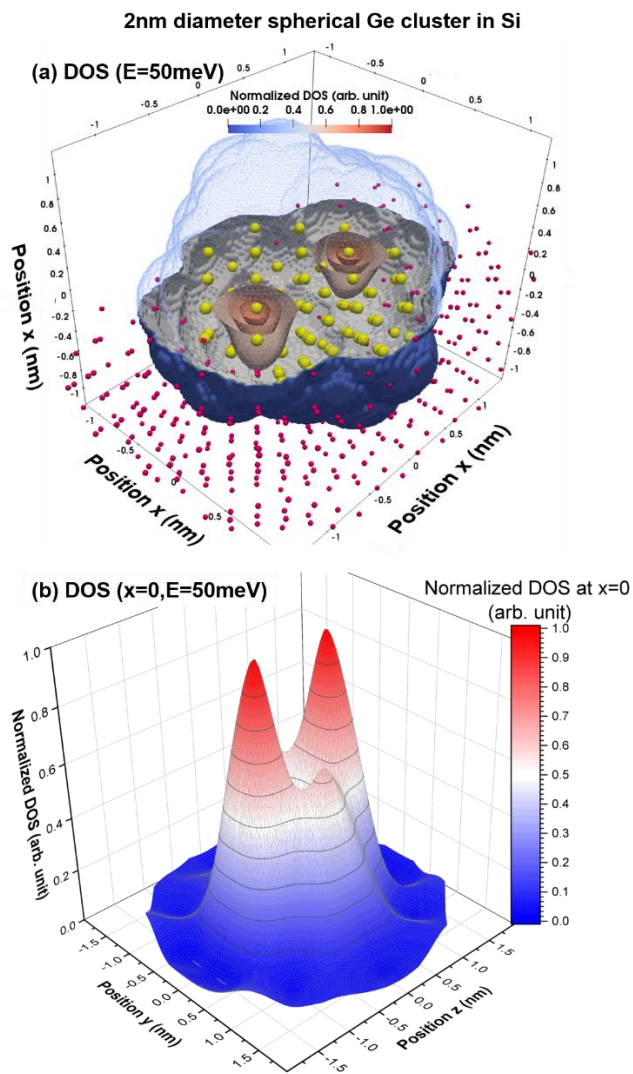

Figure 7 Normalized electronic DOS at 50meV above the band edge of Si for a 2nm spherical Ge cluster embedded in pristine Si. For better visibility, Si (purple) and Ge atoms (yellow) only up to x=0 are shown (a). The contour graph (a) and the cut of the DOS at x=0 (b) show the electronic DOS leaks significantly into the Si domain.

## ASSOCIATED CONTENT

**Supporting Information**
The structure files with the position of C and Si atoms of Figs. 4(c) and 5 can be freely downloaded from www.nanohub.org/resources/30959.

## AUTHOR INFORMATION

### Corresponding Author
* E-mail: charlesj@purdue.edu.

## Author Contributions

J.C. performed the method development and implementation, the numerical calculations, and the manuscript writing. S. K. contributed to the manuscript writing and by consultation and discussions. T. K. contributed to the method development and implementation, the data analysis and the manuscript writing. He supervised the project.

## Funding Sources

This research was supported by the NSF EFRI 2DARE 1433510 and through computational resources provided by Rosen Center for Advanced Computing at Purdue University, West Lafayette, Indiana.

## Notes

The authors declare no competing financial interest.

## ACKNOWLEDGMENT

During the preparation of this manuscript, we became aware of Ref. 64 which introduces an open system self-energy treatment for regular, pristine environments. Irregular environments such as those that form the subject in this work are beyond Ref. 64, however.

We acknowledge long and fruitful discussions with Prof. R. Graham Cooks of the Purdue Chemistry department.

## ABBREVIATIONS

NEGF, nonequilibrium Green's function method; ROBIN, Recursive open boundary and interfaces method; DOS, density of states;